# Tunable dual-comb spectrometer for mid-infrared trace gas analysis from 3 to 4.7 µm

LEONARD NITZSCHE,[1,*] JENS GOLDSCHMIDT,[1] JENS KIESSLING,[1] SEBASTIAN WOLF,[1] FRANK KÜHNEMANN,[1] JÜRGEN WÖLLENSTEIN[1,2]

[1]*Department of Gas and Process Technology, Fraunhofer Institute for Physical Measurement Techniques IPM, Georges-Köhler-Allee 301, 79110 Freiburg, Germany*
[2]*Laboratory for Gas Sensors, Department of Microsystems Engineering, University of Freiburg, Georges-Köhler-Allee 102, 79110 Freiburg, Germany*
*\*leonard.nitzsche@ipm.fraunhofer.de*

**Abstract:** Dual-frequency comb spectroscopy has emerged as a disruptive technique for measuring wide-spanning spectra with high resolution, yielding a particularly powerful technique for sensitive multi-component gas analysis. We present a spectrometer system based on dual electro-optical combs with subsequent conversion to the mid-infrared via tunable difference frequency generation, operating in the range from 3 to 4.7 µm. The simultaneously recorded bandwidth is up to 454(1) GHz and a signal-to-noise ratio of 7.3(2) x $10^2$ $Hz^{-1/2}$ can be reached. The conversion preserves the coherence of the dual-comb within 3 s measurement time. Concentration measurements of 5 ppm methane at 3.3 µm, 100 ppm nitrous oxide at 3.9 µm and a mixture of 15 ppm carbon monoxide and 5 % carbon dioxide at 4.5 µm are presented with a relative precision of 1.4 % in average after 2 s measurement time. The noise-equivalent absorbance is determined to be less than 4.6(2) ‰ $Hz^{-1/2}$.

## 1. Introduction

Established methods for mid-infrared (MIR) gas analysis are tunable laser absorption spectroscopy and Fourier-Transform infrared spectroscopy (FTIR). Approaches based on tunable lasers offer high spectral resolution and sensitivity but are often limited by their covered spectral width. FTIR systems excel in spectral coverage but struggle to provide spectral resolution below few GHz, especially in combination with high acquisition rates.

Frequency combs are likely to play an important role in bridging this gap. They typically cover a broad spectral span with evenly-spaced laser modes and are employed for various applications as distance measurements [1, 2], hyperspectral imaging [3, 4] and spectroscopy [5].

Especially the technique of dual-comb spectroscopy (DCS) [6] is a promising candidate for a novel generation of highly sensitive and fast spectrometers. There, two frequency combs with slightly different mode spacing are combined such that one comb acts as an optical clockwork [7] for the other. This results in an intensity modulation labeled as interferogram, which itself is represented by a comb in the radio-frequency (RF) domain. The spectrum can be recovered from a fast intensity measurement of the interferogram with a single photodetector and subsequent Fourier-transformation, rendering an additional spectrometer unnecessary.

Many gases of profound interest, e.g. the greenhouse gases carbon dioxide, methane, nitrous oxide, sulfur hexafluoride and toxic gases as carbon monoxide, ammonia or hydrogen sulfide - not only to humans but also poisonous for fuel cells or catalysts in power-to-gas reactors [8]. These gases exhibit strong and characteristic absorption features in the MIR. The full-width at half maximum (FWHM) of the corresponding rotational vibrational absorption lines at atmospheric pressures (1 atm) is typically few GHz or less. It requires frequency combs with sufficiently small mode spacing to sample their spectral signature, if no additional methods like frequency-sweeping are implemented [9].

Different approaches have been followed to provide dual-frequency combs, further referred to as dual-combs, in the MIR. Optical-parametric oscillators (OPO), for which an overview is given in [10], have been used for direct generation of high resolution dual combs [11, 12].

Alternatively, frequency combs from mode-locked Ti:Sa or fiber lasers [13, 14] generated in the near-infrared (NIR) with subsequent nonlinear-optical frequency conversion to the MIR can be employed [15] and allows gas sensing to analyze combustion [16] . In these realizations, the two frequency combs forming the DCS systems are generated separately. To provide the coherence between those two frequency combs, which is required for stable and coherent long-term DCS measurements, different approaches based on additional hardware and hence higher system complexity are pursued [17, 18]. In contrast, DCS systems based on two frequency combs generated by electro-optical modulation of light from a single common continuous-wave (cw) laser are mutually coherent, leading to a reduced complexity [19, 20]. Switching the mode spacing fast combined with an optical delay in one branch of an interferometer allows to even further simplify the setup [21]. The conversion of electro-optic combs into the MIR was demonstrated via intra-pulse difference frequency generation (DFG) of a single 10 GHz-comb [22], which requires multiple stages of spectral broadening, and with superimposed combs in a single DFG [23].

In this work, we follow the approach of Millot *et al.* [24] for the electro-optic dual-comb generation, which combines comb initialization via fast intensity modulation and spectral broadening in a dispersion compensating fiber. It allows for direct control of the mode spacing in the range of hundreds of MHz and spectral bandwidth of hundreds of GHz [25]. Employing a fixed-frequency pump laser for the DFG conversion yields a usable spectral range from 3.15 µm to 3.5 µm [26]. With this approach the MIR-tunability is provided by the wavelength agility of the NIR-comb itself. Limitation arises from the fact that most optical components for comb generation at 1550 nm do not provide a broad spectrum. In order to increase the tuning range of the DCS system, we combine a fixed-frequency laser source for comb generation in the NIR with a tunable optical-parametric oscillator (OPO) as pump light for the DFG to reach the MIR. Additionally, both combs are superimposed already in the NIR and converted simultaneously, allowing to maintain the intrinsic coherence and simplifying the setup.

For characterization of the spectrometer, we investigate the tunability of the system and dual-comb related quantities, including the signal-to-noise ratio (SNR), linewidth of the RF-comb modes, and spectral coverage. Using acquisition times of few seconds, the spectrometer is utilized for transmission measurements of methane, nitrous oxide and a mixture of carbon dioxide and carbon monoxide for which we determine the respective concentrations.

## 2. Experimental realization

### 2.1 Near-infrared comb-source

The DCS spectrometer is based on the generation of two frequency combs from a cw fiber laser in the telecommunications band at 1550 nm. This is achieved by fast electro-optical intensity modulators and further spectral broadening within a dispersion compensating fiber. This approach has been demonstrated by Millot et al. [24], where the setup included both combs propagating in opposite directions through the same dispersion compensating fiber. In this realization, two fully separate branches are used for comb generation to avoid any counter-propagation of optical signals before the combs fully evolve. This suppresses mixing of undesired reflections into the signals. In addition, our scheme offers flexibilities useful for system investigation, e.g. to monitor the evolution of undesired sidebands often caused by imperfections of the driving electronics.

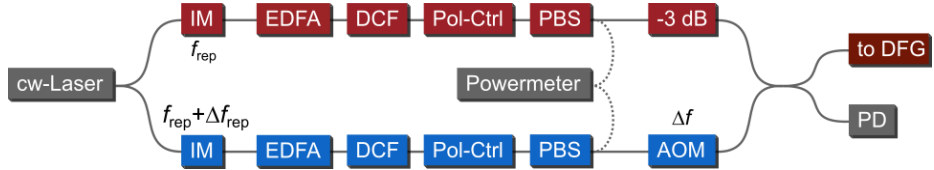

Fig. 1: Schematic of the fiber-based electro-optic dual-comb generation. Boxes indicate benchtop modules or fiber components. Solid lines stand for polarization maintaining fibers (PMF), dashed lines for PMF guiding the orthogonal polarization. All couplers have 1:1 split ratio. *Legend:* IM: Intensity modulator; EDFA: erbium-doped fiber amplifier, DCF: dispersion compensating fiber (not maintaining polarization); PBS: polarizing beam splitter; AOM: acousto-optic modulator; DFG: difference frequency generation; PD: photodetector.

The dual-comb is generated as shown in Fig. 1. A single-frequency cw laser (ADJUSTIK E15 HP / NKT Photonics) centered at 1550 nm is split into two branches with equal power. The initial combs are generated by intensity modulation with benchtop modules (Modbox-PG-CBand-50 / iXBlue), delivering pulse trains with pulse duration of roughly 50 ps and an extinction ratio larger than 40 dB. The repetition rate ($f_{rep}$) of the internal electrical pulse generator driving the intensity modulators equals the clock signals, with frequencies ranging from 250 to 500 MHz, set via an additional multi-clockboard. The difference in repetition rates ($\Delta f_{repp}$) is typically chosen in the range from 2 kHz to 20 kHz. For spectral broadening of the initial combs, spanning roughly 40 GHz, the pulse trains are amplified to average powers of 100 to 400 mW with standard cw erbium-doped fiber amplifiers (EDFA). In the second step, the combs are launched into dispersion compensating fibers (SMFDK-S-010 / OFS) where they undergo spectral broadening, and optical wave-breaking occurs [27]. Both combs are then linearly polarized by fiber-loop based polarization controllers. Polarizing beam splitters filter the desired polarization. The polarization controllers are adjusted for maximum transmission through the polarizers once after the warm-up of the EDFAs. The comb from one branch is attenuated by 3 dB where in the other branch an acousto-optic modulator (AOM) shifts one comb by 40 MHz ($\Delta f$). The electrical power driving the AOM is set such that both combs have the same average optical power. The combs are then superimposed by using a polarization-maintaining fiber-coupler, which blocks the fast axis guiding the undesired polarization. This effectively acts as an additional polarizer, further suppressing imperfections during polarization control. For driving the AOM we use a signal generator (SMA100B / R&S) which by the same time provides a reference clock. This reference clock is connected to the multi-clock board, which provides all other frequencies needed to generate the combs and to synchronize the personal computer hosted digitizer with 16-bit analog-to-digital-converters running at 250 MSPS.

*2.2 Conversion to the mid-infrared*

In order to maintain the mutual coherence of the two frequency combs upon conversion, they are superimposed and then simultaneously frequency-converted from NIR to MIR by a single DFG stage. In contrast to similar realizations based on a fixed-frequency pump laser for the DFG [23, 26], we use a tunable pump extending the tuning range of the converted MIR combs. The conversion setup is depicted in Figure 2. A continuous-wave OPO in a four-mirror bow-tie configuration acts as the pump source for the DFG. To pump the OPO we use a cw-laser with an output power of 7 W at 775 nm (Koheras HARMONIK / NKT Photonics). This laser is focused to a beam radius of 40 µm in the center of a 30 mm long periodically-poled lithium niobate crystal (PPLN) in a thermally controlled housing mounted on a linear stage. The PPLN comprises ten different channels with poling periods between 19.5 and 21.1 µm. By temperature tuning and switching between the different poling periods, the OPO emission can be tuned in a wavelength range of 1.0 to 1.3 µm, providing up to 1.0 W of output power. For frequency stabilization, the OPO is side-of-fringe-locked to a temperature-stabilized external cavity. For difference frequency generation with the dual-comb, the OPO output is coupled into a second bow-tie cavity with a finesse of 270, resonantly enhancing the power up to 70 W. This second cavity is stabilized with a Pound-Drever-Hall scheme. A 10 mm long PPLN, which has 10 different poling channels with poling periods between 27.58 and 31.59 µm, is placed in one focus of the cavity with a beam waist of 65 µm. The NIR dual-frequency comb is coupled out of the fiber and focused into the center of the DFG PPLN to guarantee for optimal overlap with the cavity mode. The DFG transfers the NIR dual-comb to the MIR in the range between 3.0 µm and 4.7 µm for an OPO-wavelength of 1.02 µm and 1.17 µm respectively. The average power of the MIR dual-comb is about 1.0 mW, which is typically attenuated to few hundreds of µW to avoid saturation of the photodetectors. After the conversion, the MIR dual-comb is collimated using a $CaF_2$ lens and split into a reference and a sample branch with a pellicle beam splitter. The reference branch is guided directly to a HgCdTe-detector. In the sample branch, the dual-comb propagates through a Herriott-type multi-reflection cell with an optical path length of 720 cm and is detected with a second HgCdTe-detector. Both detectors (PVI-4TE / Vigo) exhibit a high detection bandwidth (> 100 MHz).

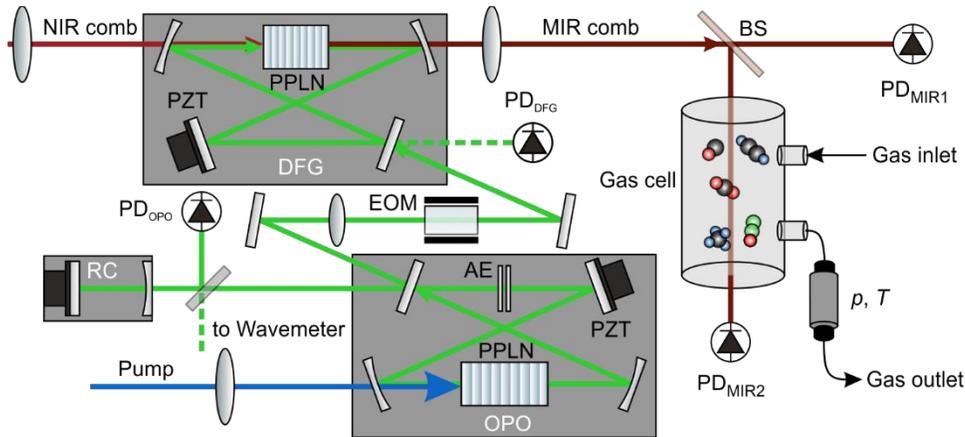

Fig. 2. Sketch of the dual-frequency-comb converter. The near-infrared (NIR) combs are converted to the mid-infrared (MIR) by difference frequency generation (DFG) in a bowtie-cavity. The continuous-wave pump light for the DFG is generated by an optical-parametric oscillator (OPO) in a second bowtie-cavity, which is stabilized to an external cavity (RC). For that, a small fraction of the OPO light reflected from a surface of the air-spaced etalon (AE) is used. In addition a part of this reflected light is fed to a wavelength meter. The OPO itself is pumped with a continuous-wave laser at 775 nm. For gas analysis, the MIR dual-comb is split into two branches where one branch is detected with a fast photodiode ($PD_{MIR1}$) while the other propagates through a multi-reflection flow cell with 720 cm absorption path before detection by $PD_{MIR2}$. The pressure ($p$) and temperature ($T$) of the gas sample are measured downstream of the cell. For clarity the locking electronics, for which $PD_{OPO}$ and $PD_{DFG}$ provide the signals, are not depicted. *Legend:* (BS) beam splitter, (EOM) electro-optical modulator, (PZT) piezo-mounted mirror, (PPLN) periodically poled lithium niobate, (BS) beam splitter.

## 3. Mid-infrared dual-frequency comb performance

As the spectrometer performance is predominantly determined by the quality of the frequency comb, we first focus on basic comb quantities like the signal-to-noise ratio (comb SNR) and the detectable number of comb modes *M*, which yields the spectral span when multiplied with the repetition rate. In addition, we determine the linewidth of the radio-frequency beat notes, which gives insights on the coherence of the MIR combs.

For that we define two configurations making use of the flexibilities in repetition rates and spectral bandwidth of the comb-source. The first configuration is conducted with $f_{rep}$ = 250 MHz, where $\Delta f_{rep}$ is set to 5 kHz and the average output power for both EDFAs is 400 mW. The other configuration is using $f_{rep}$ = 500 MHz, $\Delta f_{rep}$ = 20 kHz, and 100 mW EDFA output power. The higher EDFA power for the 250 MHz configuration leads to higher pulse peak powers for increased spectral broadening of the combs in the DCF. For both configurations we record time series of 1 s total measurement time across the full tuning range, and calculate the respective amplitude spectra. Linear detection is ensured by carefully reducing the power levels on the photodetector until out-of-band signals in the radio-frequency spectrum are eliminated [28]. For both configurations, we use a fixed detection bandwidth of 10 MHz centered at 40 MHz. Four dual-comb spectra with wavelengths of 1550 nm, 3000 nm, 3900 nm and 4700 nm are plotted in Fig. 3 (right) for the 250 MHz configuration.

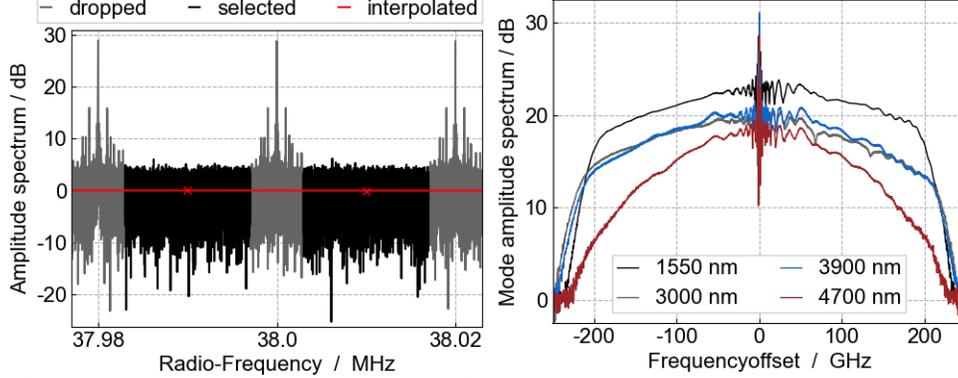

Fig. 3: (left) Zoom in on the amplitude spectrum (1 Hz resolution) showing the resolved comb modes at 38 MHz ± 20 kHz. Sidebands originating from imperfections in electronics and vibrations are visible in the 6 kHz band – grey – centered at each mode. To derive the noise floor level for the comb modes, we average the noise floor amplitudes between the comb modes – black – and interpolate them – red line. (right) Dual-comb envelopes retrieved from the amplitude spectra of a single trace of interferograms acquired over 1 s for 250 MHz repetition rate. All amplitudes are corrected by the minimal value of the averaged interpolated noise floor – red line (left) – for convenience. For higher wavelengths, the peak conversion efficiency of the DFG as well as the FWHM of the conversion gain bandwidth decrease, which is clearly visible in the 4700 nm spectrum.

We compute the SNR by taking the average over all amplitudes of all radio-frequency comb modes higher than 10 dB with respect to the noise floor in-between the comb modes, such that the comb modes consistently exceed the standard deviation of the surrounding noise. The noise level at the frequencies of comb modes is estimated by interpolating the average amplitude of the noise floor between the comb modes where no sidebands are present as highlighted by black areas shown in Fig. 3 (left). The amplitude of the comb modes is given by the maxima in the amplitude spectra – here only shown exemplary for 37.98, 38.0 and 38.02 MHz modes. This evaluation also yields the effective number of comb modes $M$. The comb mode sidebands visible in the RF spectrum are proportional to the amplitude of their adjacent comb mode. The dominating sidebands are located at a 1 kHz offset with approx. -15 dB with respect to the comb mode. They originate from the active modulation used to stabilize the intensity modulators for comb generation. Other sidebands are attributed to mechanical noise, e.g. from a compressor for water cooling of our EDFAs. Those can couple into the DCF, leading to polarization state modulations and, in conjunction with the polarizers, to intensity modulations. Consequently, operating at $\Delta f_{rep} > 4$ kHz avoids any interference by parasitic sidebands.

Table 1. Dual-Frequency comb characteristics in the near- and mid-infrared

| Mode spacing | 250 MHz | 500 MHz | 250 MHz | 500 MHz |
|---|---|---|---|---|
| Central wavelength | Number of comb modes | | SNR [Hz$^{-1/2}$] | |
| | Spectral coverage [GHz] | | Figure of merit [Hz$^{-1/2}$] | |
| 3000 nm | 1902(6) | 488(3) | 1.7(1) x 10$^2$ | 6.6(9) x 10$^2$ |
| | 475(1) | 244(1) | 3.3(3) x 10$^5$ | 3.2(4) x 10$^5$ |
| 3900 nm | 1861(10) | 486(1) | 1.7(2) x 10$^2$ | 7.3(2) x 10$^2$ |
| | 465(2) | 243(1) | 3.1(4) x 10$^5$ | 3.5(1) x 10$^5$ |
| 4700 nm | 1661(22) | 469(2) | 1.0(1) x 10$^2$ | 4.8(5) x 10$^2$ |
| | 415(6) | 234(1) | 1.6(2) x 10$^5$ | 2.2(2) x 10$^5$ |

The dual-comb characteristics in the MIR evaluated at the central wavelength of 3000, 3900 and 4700 nm for both configurations are listed in Tab. 1. Switching between two arbitrary central wavelengths is determined by the time required for the tuning of the OPO. In the current setup, this requires manual adjustments and takes several minutes. The SNR is consistently larger than $10^2\,\text{Hz}^{-1/2}$ and a maximal value of $7.3(2) \times 10^2\,\text{Hz}^{-1/2}$ is reached for 500 MHz repetition rate at 3900 nm. The figure of merit [29] defined as SNR times $M$ is in the range of $10^5\,\text{Hz}^{-1/2}$ for both configurations, which is similar to other dual-comb realizations [6]. The comb spectra are impacted by the DFG towards longer MIR wavelengths via two major mechanisms. First, the conversion window given by the phase-matching condition decreases towards longer target wavelengths due to dispersion of the PPLN. Second, the onset of absorption in the PPLN above 4.5 µm decreases the overall power of the generated DFG combs, thus reducing the number of lines above the 10 dB threshold criterion. Both effects constrain the spectral coverage of the combs towards the long-wavelength end. The impact of the DFG conversion becomes apparent in the spectrum at 4700 nm, where it leads to a reduced average SNR of the comb amplitudes – especially for the 250 MHz configuration. This is also visible in the comb mode spectrum plotted in Fig. 3 (right). The full bandwidth at -20 dB of the DFG conversion gain is roughly 440 GHz for the conversion to 4700 nm according to simulation [30] using absorption data from [31] and refractive indices from [32]. For a conversion to 3900 nm the full bandwidth at -20 dB exceeds the maximum span of the dual-combs by far, and thus imposes no narrowing effect on the comb span.

The evolution of the SNR depending on the integration time is of particular interest when aiming for high spectra acquisition rates – a key advantage of dual-comb-based spectrometers. For the analysis the amplitude spectra for different lengths of the time series of interferograms taken at 3 µm wavelength are computed and the SNR is determined with the procedure presented above. The mode threshold is set to 0 dB instead of 10 dB to account for lower amplitudes occurring for short integration times. The resulting SNR is plotted in Fig. 4 (left) for the average over all comb modes for both the 250 MHz and 500 MHz configurations (all). In addition, the evolution of a weak mode (500 MHz setup) at 35.5 MHz in the RF-domain, which is situated at the edge of the comb spectrum, and a high power comb mode at 39.5 MHz (strong) close to the center frequency are plotted.

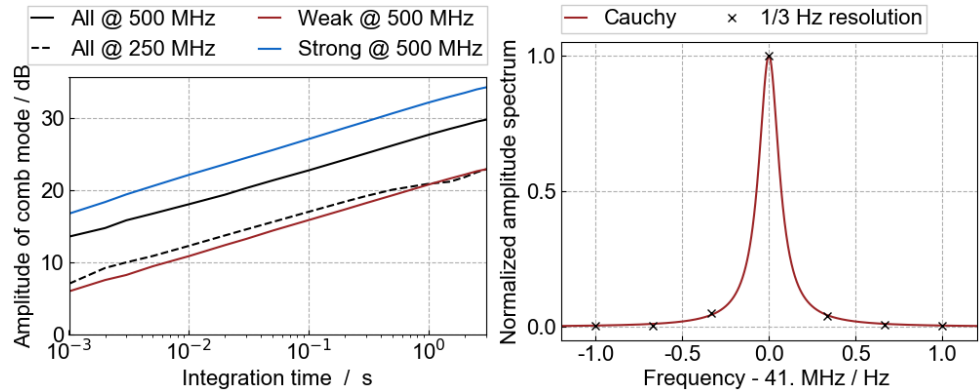

Fig. 4. (left) Amplitude of the comb modes calculated from a continuously digitized interferogram series with repetition rates of 500 MHz or 250 MHz. All represents the average over all comb modes for the respective repetition rates. In addition the evolution of a mode residing at the edge of the comb spectrum (weak) and one in the center (strong) are plotted. The interferogram series is cropped to the respective integration time. (right) Zoom-in on a single comb mode evaluated for 3 s time series which yields a frequency resolution of 0.33 Hz. The 3 s spectrum is used to fit a Cauchy distribution, which results in a FWHM of 0.14 Hz but is most likely not sufficiently resolved.

As expected the evolution of the weak and strong mode amplitudes scale with the square root of integration time. Setting the mode threshold to 0 dB does not allow to discriminate between real comb modes and the noise floor at the edges of the comb spectrum. Hence the evolution of the averaged mode amplitudes exhibit deviations from this scaling – especially at shorter integrations times. However, the overall scaling is still feasible. For a repetition rate of 250 MHz more comb modes are generated, effectively distributing the total detectable optical power over more amplitudes and hence the average mode amplitude is reduced compared to the 500 MHz configuration. To quantify the coherence of the combs, we evaluate the comb mode linewidth for 3 s integration time. In Fig. 4 (right) the normalized spectrum zoomed to one comb mode is shown for the resulting 0.33 Hz resolution exemplarily. Fitting a Cauchy distribution to the data results in a FWHM of 0.14 Hz, which is below the resolution limit. This highlights the advantage of the mutual coherence possible using a common laser for comb generation. This property is beneficial when designing the algorithms for spectral computation and renders any correction or active stabilization unnecessary.

From both the evolution and linewidth analysis of the dual-comb mode, we conclude that the coherence of the combs reaches up to 3 s. The optical linewidth of a single comb mode cannot be determined with this procedure, but derived from the OPO light and cw-laser used to generate the combs. It is estimated to be in the order of single digit MHz or less. The major contribution comes from the OPO, which we know from other realizations in our labs. The contribution from the cw-laser can be neglected as it is below 1 kHz according to the manufacturer. Other contributions caused by phase noise from the modulators and fibers in both comb branches – especially the 1 km long DCF - during comb generation have a negligible effect as well according to the presented linewidth analysis.

### 4. Retrieval of transmission spectra

The basic data for the determination of species concentrations in a gas sample are transmission spectra derived from the measurements. In the following, the steps to retrieve transmission spectra are described. In addition, we investigate the validity of this procedure and which corrections can further be undertaken to compensate for imperfection.

In general, amplitude spectra are computed via FFT from time traces of one second duration. As we aim for high sensitivity in gas analysis, we use the 500 MHz configuration in the following to utilize the highest comb SNR possible. The signals of the reference and sample detector are recorded simultaneously and labeled as $S_{\text{ref}}$ and $S_{gas}$. From these amplitude spectra, the maxima are selected, and their ratio spectrum $RS_{gas}$ accounts for the comb envelope structure. The ratio spectrum is given by

$$RS_{gas} = S_{gas}/S_{\text{ref}}. \quad (1)$$

From these ratio spectra the transmission spectra $T$ are calculated as the ratio of subsequent measurements from both a sample gas and nitrogen ($N_2$) filled cell in order to account for the cell's transmission properties by again calculating their ratio according to

$$T = RS_{sample}/RS_{\text{N2}}. \quad (2)$$

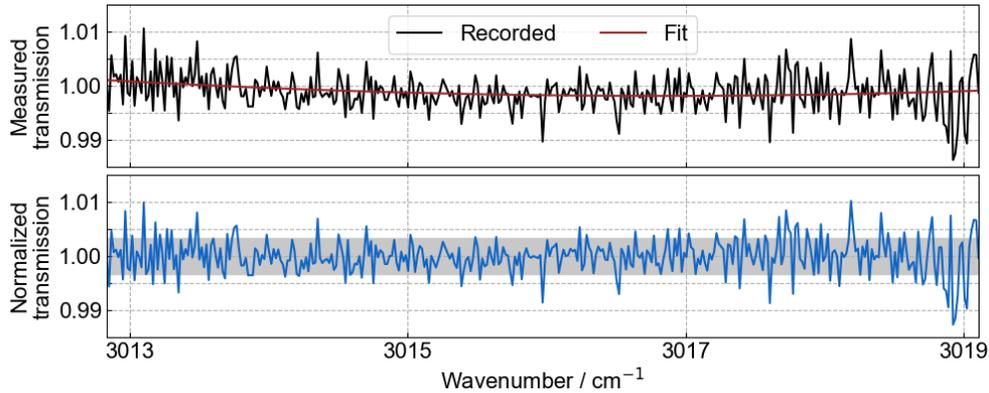

Fig. 5. Comparison of two nitrogen spectra taken at 3.3 µm. The recorded data represent the ratio of two nitrogen spectra derived from 1 s acquisition time and taken 3 minutes apart. A 2$^{nd}$-order polynomial function is fitted to the recorded spectrum and used for normalization. The RMS-noise of the normalized spectrum for this measurement is 3.2 x 10$^{-3}$ indicated by the grey shaded area.

An exemplary "100-%" transmission spectrum is shown in Fig. 5, where the $N_2$-filled cell is taken for both, sample and the $N_2$ measurement - again acquired in 1 s each. Both spectra are taken 3 minutes apart. When repeating these type of measurements we typically find systematic deviations from the expected 100-% transmission. This originates from instabilities in the comb envelopes, which are most likely caused by drifts in optical powers in between the sample and reference measurements. We treat these deviations, in the following referred to as baseline, by fitting a 2$^{nd}$-order polynomial function ($\text{poly}^{(2)}$). The fit results are then used to normalize the derived transmission spectra, yielding the normalized transmission shown in Fig. 5. This procedure has to be repeated for each measurement and will be an integral part of the fit function to determine gas concentrations later. The noise equivalent absorbance (NEA) calculated from the RMS-noise of the normalized transmission spectrum, for which 2 s measurement time were used in total, is 4.6(2) x 10$^{-3}$ Hz$^{-1/2}$. The uncertainty represents the standard deviation of the NEA for 6 measurements. Note that we consider the integration times of both nitrogen spectra, which are 1 s each, to normalize with respect to the bandwidth. The use of higher order polynomials does not further improve the NEA. Hence it is chosen as hypothesis for the baseline underlying the transmission of a sample. The low polynomial order of the baseline is beneficial as it will be more easily discriminated from more complex structures, e.g. line profiles of gas absorptions, during fitting later.

The x-axis is calculated for each NIR comb separately, where the optical mode spacing is $f_\text{rep}$ ($+\Delta f_\text{rep}$ likewise) and one comb is shifted by $\Delta f_\text{rep}$. The repetition rate $f_\text{rep}$ is derived from the master clock (internal reference clock of R&S SMA100B) with an accuracy better than 10$^{-7}$. The axes are averaged and shifted to the central spectral frequency in the MIR, where the central frequency is estimated by the difference frequency between the OPO light and the NIR cw-laser used for comb generation. The frequency of the free-running NIR cw-laser is known with a precision of 5 GHz while short-term drifts are 0.01 MHz/s in the worst case known from an a-priori calibration, for which a wavelength meter (WS7 / High Finesse) was used. The frequency of the OPO is measured continuously during the operation of the spectrometer and drifts of up to 1 MHz/s are observed. This motivates to limit the data acquisition to few seconds to avoid smearing out of the absorption profiles laser. Drifts of the NIR cw-laser do not contribute significantly. The remaining uncertainty in the absolute x-axis offset is accounted for by allowing a global frequency shift $x_0$ during fitting of the spectra.

As a side effect of the system layout with superimposed combs, both combs interact with the sample, but at slightly different optical frequencies, in average $\Delta f = 40$ MHz apart. While the resulting beat note represents a single spectral element, it contains the attenuation information from two slightly different spectral positions, leading to systematic deviations of the retrieved data from the true spectrum. The magnitude of the deviation caused by the 40 MHz split depends on the widths of the absorption lines. As an example, the resulting absorption deviations for nitrous oxide at 3.9 µm with 100 ppm concentration under atmospheric pressure are calculated to be less than 2 ‰. Compared with the NEA obtainable with 2 s acquisition time (see Fig. 5) these effects remain too small to be observed.

## 5. Analysis of trace gases

To calculate the concentrations of gas species under investigation for a given sample, we fit a model function $f_{\text{fit}}$ to the retrieved spectra. It is composed of the baseline hypotheses and a simulated transmission spectrum using the HITRAN database [33], which composes all absorptions lines simulated with Voigt-profiles within the measured spectral range. For compatibility with HITRAN, the values of the x-axis are converted to wavenumbers [cm$^{-1}$] before fitting:

$$f_{\text{fit}}(x; x_0, \overrightarrow{VMR}, \vec{c}\,) = \exp\left[-\alpha\left((x-x_0); \overrightarrow{VMR}\right) * L\right] * \text{poly}^{(2)}\left((x-x_0); \vec{c}\,\right). \quad (3)$$

Free parameters are a global wavenumber offset $x_0$, the volume mixing ratios ($\overrightarrow{VMR}$) of the gas species used to calculate the absorption coefficient α and the baseline coefficients $\vec{c}$. The total absorption path length $L$ is 720 cm. Pressure and temperature of the gas cell are held constant at their respective measured values.

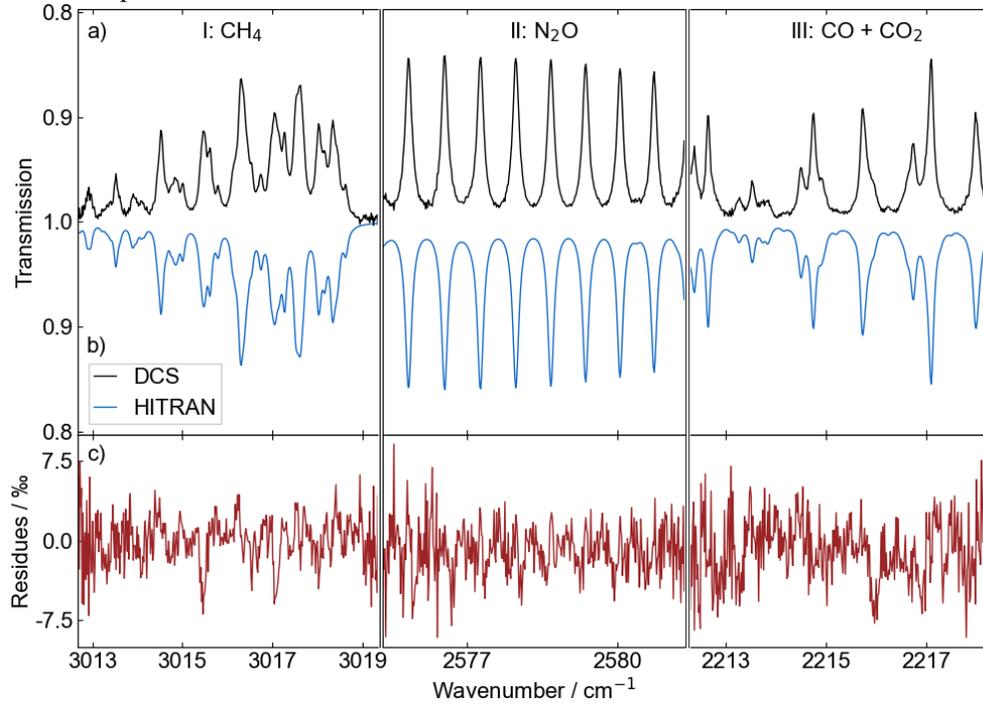

Fig. 6. a) Transmission spectra of methane (CH$_4$) (I), nitrous oxide (N$_2$O) (II) and a mixture of carbon monoxide (CO) with carbon dioxide (CO$_2$) (III) at atmospheric pressures and room temperature. For the individual measurements I to III, the central wavelength of the dual frequency comb was set to 3.3, 3.9 and 4.5 µm, respectively. b) Simulated spectra based on HITRAN providing the best fit to the experimental spectra. c) Residues between measured spectra and model fit. The residues exhibit deviations with RMS-noise below 3.2 x 10$^{-3}$. Systematic deviations are expected as the used HITRAN parameters, e.g. pressure broadening coefficients are stated to be correct only within 2 % to 5 % for N$_2$O at 3.9 µm, and contribute to the residues as well.

The versatility of the spectrometer is demonstrated by three separate measurements of CH$_4$ at 3.3 µm, N$_2$O at 3.9 µm and a mixture of CO and CO$_2$ at 4.5 µm, denoted in the following as measurements I, II and III. With those, we cover different parts of the accessible spectral range while the different lineshapes and overlaps of the gas absorptions provide a variety of scenarios to test the fit procedure to the respective transmission spectra according to eq. 3. Again, nitrogen spectra are taken first, and transmission spectra are derived according to eq. 2. The measured transmission spectra as well as simulations based on the fit results are shown in Fig. 6. For convenience, the plotted transmission spectra are already normalized with the baseline poly$^{(2)}$ using the coefficients $\vec{c}$ determined by the respective fits. For I, a mixture of 103.3 ppm ± 2 % $CH_4$ in synthetic air is further diluted with nitrogen to reach a target concentration of 5 ppm. For, II the supplied mixture of 99.8 ppm ± 2 % $N_2O$ in nitrogen was used without further dilution. For measurement III, we mixed two supplied dilutions of 59.5 ppm ± 2 % CO in synthetic air and 19.9 % ± 2 % CO2 in synthetic air with 50:50 ratio. This mixture was further diluted with $N_2$ to further reduce the concentrations to approximately 15 ppm CO and 5 % $CO_2$. For mixing of the gases, three mass flow controllers (EL-FLOWprestige / Bronkhorst) were used. The used gas mixtures, the respective target gas concentrations with their uncertainties derived from gas mixtures and mixing procedure, and the concentrations obtained by the model fit are listed in Table 2. In order to calculate the lineshapes, the air-broadening coefficients from the HITRAN database were used.

**Table 2. Set and measured gas concentrations for measurements I, II and III**

|   | Central wavelength and wavenumber | Sample: Gas / Diluent | Expected concentration | Derived concentration ($\sigma_{fit}$ only) | Derived concentration (incl. $\sigma_{+hitran}$) |
|---|---|---|---|---|---|
| I | 3.3 µm<br>3016 cm$^{-1}$ | $CH_4$ / air + $N_2$ | 5.0(1) ppm | 5.16(2) ppm | 5.2(1) ppm |
| II | 3.9 µm<br>2578 cm$^{-1}$ | $N_2O$ / $N_2$ | 99(2) ppm | 95.8(2) ppm | 96(2) ppm |
| III | 4.5 µm<br>2215 cm$^{-1}$ | CO / air + $N_2$ | 14.8(3) ppm | 15.3(1) ppm | 15.3(1) ppm |
|   |   | $CO_2$ / air + $N_2$ | 4.9(1) % | 4.58(2) % | 4.58(6) % |

The uncertainties on the derived concentrations represent either the uncertainties of the fit results $\sigma_{fit}$ or in addition the uncertainties of the line intensities $\sigma_{+hitran}$ according to HITRAN, which are between 2 % and 5 % for $CH_4$ and $N_2O$, 1 % and 2 % for $CO_2$ and below 1 % for CO. From those uncertainties, we picked the minimal values of all lines listed in the evaluated spectral range - typically corresponding to the most intense lines - to estimate the uncertainty $\sigma_{+hitran}$. For $CH_4$, $N_2O$ and CO, the derived concentrations are within 2$\sigma$ of the expected concentration, for $CO_2$ the derived concentration is within 4$\sigma$. The noise-equivalent-absorbance calculated from the RMS-noise of the residues is 4.5 ‰ Hz$^{-1/2}$ of measurement time considering both acquisition times for the nitrogen and sample spectrum of 1 s each. It matches with the findings for the nitrogen spectra comparison presented before.

The average precision of the determined concentrations considering the uncertainties $\sigma_{+hitran}$ is 1.4 %. Compared to the deviations from the expected concentrations indicates that the fit procedure and system has to be further developed and investigated in terms of accuracy. More precisely, the assumption that the lineshapes can be estimated using air-broadening coefficients may not hold and translate to an incorrect baseline estimation by the fit. This effectively reduces the area under the absorptions, which scales linearly with the determined concentration. The largest deviation found for $CO_2$ in measurement III can be explained by the fact that the temperature sensor was placed downstream of the gas outlet of the cell and outside a housing covering the resonators and cell. This might introduce a systematic offset of the measured temperature and effect the absorptions of $CO_2$ significantly as the lines in this spectral region are mostly hot bands exhibiting a strong temperature dependence. Further developments and investigations will require to use calibrated gas mixtures – best diluted in synthetic air only to ensure the needed parameters are provided by the HITRAN database. The temperature and pressure sensors could be directly installed inside the transmission cell. Further improvements in system stability might allow to improve the photometric accuracy to a level that the baseline impact cannot be resolved and hence can be excluded from the fit.

## 6. Conclusion

We present a mutually coherent dual-comb spectrometer for gas sensing, freely tunable over the mid-infrared range of 3 to 4.7 µm. For this purpose, an EOM-based dual-comb is converted from the NIR via DFG to the MIR. Using a tunable OPO as pump allows to position the dual-comb across the full range of 3 to 4.7 µm, where a single spectrum spans 450 GHz (15 cm$^{-1}$) for 250 MHz repetition rate and 240 GHz (8 cm$^{-1}$) for 500 MHz. The figure of merit is in the range of $10^5$ Hz$^{-1/2}$ across the full tuning range, which is consistent with other comb implementations and a factor of 2 lower in the MIR when compared to the NIR. The coherence time of the combs derived from the FWHM of the rf-comb modes is 3 s. Transmission spectra of $CH_4$ at 3.3 µm, $N_2O$ at 3.9 µm and a mixture of CO and $CO_2$ at 4.5 µm are taken in 3 different measurements and show good agreement with simulations using the HITRAN database while the normalized NEA is consistently below 4.6(2) ‰ Hz$^{-1/2}$. The concentrations of the gases are determined with an average relative uncertainty of 1.4 % in 2 s of total integration time. This validates the potential of the technique of dual-comb spectroscopy – the parallel sensing of multiple spectral elements to cover multiple absorptions features while maintaining high sensitivity and acquisition speed.

Future challenges are to improve the robustness, especially for the NIR-to-MIR conversion, in order to further improve on SNR and to allow for a field-deployable system. In addition, converting to longer wavelengths around 10 µm would allow to address a different set of gas species.

## Acknowledgements

This work was supported by the Fraunhofer and Max Planck cooperation programme.
The authors express their thanks to the group of G. Millot for the inspiring discussions on EOM-based frequency combs, G. Rieker for his comments regarding this paper and N. Picqué for the introduction to and advice on dual-comb spectroscopy.